# Chaos assisted decay of $^{180}Ta^m$

N. Auerbach
School of Physics and Astronomy
Tel Aviv University
Tel Aviv, 69978
Israel

Abstract: The decay of the J=9$^-$ isomeric state in $^{180}Ta$ is described in a model based on the phenomenon of chaos assisted tunneling. The model attempts to explain the enhanced decay of the isomeric state when irradiated by gamma-rays or subjected to Coulomb excitations.

The $^{180}Ta$ nucleus is the rarest naturally occurring isotope. It exists in an isomeric state with a half-life time of $t_{1/2} \geq 1.2 \times 10^{15}$ years at an excitation energy of 75 keV and spin $J = 9^-$ [1-4]. The ground state is a $J = 1^+$ and its half-life time is 8.1 hours. Because of the large spin gap between the ground state (g.s.) and the isomeric state, electromagnetic transitions between these two states are strongly suppressed. Moreover, the $J = 1^+$ g.s. is believed to belong to a K=1 band while the isomeric is the basis of a K=9 band. The K quantum number at these low-energies is conserved and a transition between K=1 and K=9 states is hindered.
The g.s. $K = 1$ band is a result of a collective rotation of the nucleus around the axis perpendicular to the symmetry axis, while the higher $K$ band is due to a non-collective rotation of only few nucleons around the symmetry axis. The transition between such two rotational schemes is hindered because of the very different nature of these two rotations and large rearrangement is required in order to go from one form of motion to the other. At higher energies however $K$-mixing may occur and various models of such mixing have been proposed.
When a $\gamma$-ray (or virtual $\gamma$ in Coulomb excitation) of multipolarity $\lambda$ interacts with $^{180}Ta^m$ it will excite states with $K'= |K \pm \lambda|$. For low multipolarities ($\lambda \leq 2$) excited in $^{180}Ta$ will have values of $K'$ still to high to have appreciable decay rates to the $K = 1$ g.s.
One can ask the question, whether exciting the isomer to a higher state may induce a decay that will lead to the population of the ground state. Several experiments in the recent years have provided an answer to this question. The isomer was excited by various means to intermediate states which than emitted radiation and populated in addition to the isomeric level also the g.s. . This process was observed for intermediate levels having energies as low as 1 MeV above the g.s. [5].
These findings have some important implications on questions concerning the nucleosynthesis of $^{180}Ta$ [6]. One major problem however in understanding the results of depopulation of the isomeric state via an intermediate state, is the large discrepancy between the experimental results and theoretical calculations. The rates of decay to the g.s. found in calculations are two to three orders of magnitude smaller than the measured ones [7].
In the present note we describe a mechanism that might contribute to the enhancement of the decay rates from the intermediate states to the ground level.

The two shapes of the nucleus characterizing the two bands, described above, maybe represented by two potentials in some collective coordinate, separated by a barrier. The mechanism we discuss may produce strong K-mixing due to enhancement of quantum tunneling from one shape to the other when there is an onset of chaos in nuclear dynamics [8-10].

The unperturbed states belonging to the deeper potential (the one giving rise to the $K=1$ band) will be denoted using roman letters ($|a\rangle$, $|b\rangle$, etc), the unperturbed states in the second well (the ones giving rise to the $K=9$ band) will be assigned Greek letters ($|\alpha\rangle$, $|\beta\rangle$, etc).

A state $|\alpha\rangle$ belonging to the second potential well might mix with one (or a few) special states belonging to the first potential well. These will be "doorways" through which states in the second well will couple to states in the first well.

For the sake of simplicity we limit ourselves to a single doorway, denoted $|d\rangle$. This doorway is not necessarily close in energy to the state $|\alpha\rangle$, to which it couples. Denote, $\Delta E = E_\alpha - E_d$. Using perturbation theory the admixed state will be:

$$|\tilde{\alpha}\rangle = |\alpha\rangle + c_{\alpha d}|d\rangle, \qquad (1)$$

where:

$$c_{\alpha d} = \frac{\langle \alpha | V_t | d \rangle}{\Delta E}, \qquad (2)$$

$V_t$ is an interaction that couples the state $|\alpha\rangle$ to the doorway. The coefficient usually obeys, $c_{\alpha d}^2 \ll 1$.

The doorway state $|d\rangle$ is not an eigenstate, even when one considers states in the first well only. The two-body residual interaction will mix the various unperturbed states in the first potential well, forming states $|q\rangle$. Singling out the $|d\rangle$ state any $|q\rangle$ state can be written as:

$$|q\rangle = x_{dq}|d\rangle + \sum_{a'} x_{a'q}|a'\rangle \qquad (3)$$

where $|a'\rangle$ are all other states besides $|d\rangle$ belonging to the first well.

For a strong residual interaction the doorway is spread over many states. When the mixing is complete, that is when the doorway is uniformly spread over an energy interval of the order of $\Delta E$.

The average of the coefficient $|x_{dq}|$ will be of the order:

$$\langle |x_{dq}| \rangle = \frac{1}{\sqrt{N}} \qquad (4)$$

where N is the number of states $|a\rangle$ in the interval $\Delta E$ (thus N is proportional to the density of states). The mixing between the state $|\alpha\rangle$ and the closest in energy state $|q\rangle$, will result in a state:

$$|\tilde{\alpha}\rangle_q = |\alpha\rangle + c_{\alpha q}|q\rangle \qquad (5)$$

where:

$$c_{\alpha q} = \frac{\langle \alpha|V|q\rangle}{E_\alpha - E_q} \quad \text{with} \quad E_\alpha - E_q \approx \frac{\Delta E}{N} \qquad (6)$$

Assuming that $|\alpha\rangle$ couples only to $|d\rangle$ we may write:

$$c_{\alpha q}^2 = x_{dq}^2 \langle \alpha|V_t|d\rangle^2 \frac{N^2}{(\Delta E)^2} \qquad (7)$$

which becomes:

$$c_{\alpha q}^2 = \frac{\langle \alpha|V_t|d\rangle^2 N}{(\Delta E)^2} \equiv N c_{\alpha d}^2 \qquad (8)$$

Hence, in the limit of chaos there is a large enhancement in mixing between the two K bands. The mixing is proportional to the density of states.
In studies of parity mixing in the compound nuclear states in heavy nuclei one finds enhancements of the mixing coefficient to be of the order of $10^3$ (meaning that the probabilities are enhanced by a factor $10^6$) compared to the single-particle case [11-13].
Aberg studied [9] the mixing of deformed and super-deformed bands, and found that in $^{194}Pb$ the onset of chaos occurs at excitation energies of about 2.5 MeV and the enhancement in the tunneling probability (equivalent in our case to $c_{\alpha q}^2$) is of the order of $10^2 - 10^3$ at this energy.
The above considerations are based on perturbation theory arguments using the fact that that the matrix element $\langle \alpha|V|a\rangle$ are very small compared to $\Delta E$ or even $\Delta E/N$.
It has been demonstrated however that the enhancement in tunneling, in the chaotic limit, occurs also when full diagonalization is performed and is not merely a result of the use of perturbation theory.

We may think about the following mechanism for the decay of $^{180}Ta^m$ when it subjected to $\gamma$-ray radiation, real or virtual (in Coulomb excitation). The $\gamma$-ray of low multipolarity ($\lambda = 1,2,3$) causes a transition from the isomeric state to an $|\alpha\rangle$ state with $K' = K - \lambda$ in the second well. If this $|\alpha\rangle$ state is at a sufficiently high energy, assisted by the by chaos existing in the $|a\rangle$ states in the first well, it will quickly tunnel into the first well in spite of the fact that the $|a\rangle$ states have lower K-values. In the next step the nucleus will emit $\gamma$-rays of low multipolarity decaying (either directly or by cascade) to the K=1 ground state band.

Experiments show that the depopulation of the isomer occurs already for intermediate states with excitation energies of about 1 MeV excitation. As the energy
of the photons exciting the isomer increases and higher intermediate states are reached, the rate of depopulation grows very fast. When the intermediate state is around 2.8 MeV the rate of decay is two orders of magnitude higher than for the 1 MeV intermediate state.

The most detailed theoretical studies of the Ta isomeric state were performed by Soloviev [7] using quasi-particle+phonon states in a deformed basis. The main findings of this work is that there no decay of the isomer to the g.s. for intermediate states below 2.3 MeV excitation. The depopulations sets on only at energies above it, but even then the rates are two orders of magnitude smaller than experiment. This should not be surprising in view of the above comment. As Soloviev indicates himself, the space he uses is two small to reach the chaotic stage. An extension to multi quasi-particle and multi phonon states are required to reach chaos, but that is beyond the scope of Soloviev's calculations.

It is doubtful whether the mechanism discussed in the present work can contribute to the decay from the lowest intermediate state at 1 MeV. However, also experiments indicates a very slow depopulation at this energy. As one goes to higher excitation energies, E>2.5 MeV, the depopulation experimentally increases by two orders of magnitude. As one reaches these energies the nucleus enters into the chaotic regime and the mechanism described here should produce enhancements of the same magnitude.

## Acknowledgements

We wish to thank B. Barrett, L. Zamick for discussions, and B.Jennings for his hospitality at TRIUMF where part of this work was performed.